\documentclass[10pt,aps,prd,twocolumn,showpacs,nofootinbib,preprintnumbers]{revtex4-1}
\pdfoutput=1
\usepackage{fullpage}
\usepackage{amsmath}
\usepackage{subfigure}
\usepackage[pdftex]{graphicx}
\usepackage[english]{babel}
\usepackage{color}
\usepackage{dcolumn}
\usepackage[colorlinks]{hyperref}
\usepackage{cancel}
\usepackage{units}
\usepackage[normalem]{ulem}
\hypersetup{
pdfauthor={},
pdftitle={Higgs Mass and Planck Scale}
}

\begin{document}
\allowdisplaybreaks[1]
\widetext

\title{Planck Scale Boundary Conditions and the Higgs Mass}
\vspace*{10mm}
\author{Martin Holthausen}\email{martin.holthausen@mpi-hd.mpg.de}
\affiliation{Max Planck Institute for Nuclear Physics,Saupfercheckweg 1, 69117 Heidelberg, Germany}
\author{Kher Sham Lim}\email{khersham.lim@mpi-hd.mpg.de}
\affiliation{Max Planck Institute for Nuclear Physics,Saupfercheckweg 1, 69117 Heidelberg, Germany}
\author{Manfred Lindner}\email{lindner@mpi-hd.mpg.de}
\affiliation{Max Planck Institute for Nuclear Physics,Saupfercheckweg 1, 69117 Heidelberg, Germany}
\date{\today}
\begin{abstract}
If the LHC does only find a Higgs boson in the low mass region and no other new physics, then one should reconsider scenarios where the Standard Model with three right-handed neutrinos is valid up to Planck scale. We assume in this spirit that the Standard Model couplings are remnants of quantum gravity which implies certain generic boundary conditions for the Higgs quartic coupling at Planck scale. This leads to Higgs mass predictions at the electroweak scale via renormalization group equations. We find that several physically well motivated conditions yield a range of Higgs masses from \(\unit[127-142]{GeV}\). We also argue that a random quartic Higgs coupling at the Planck scale favours \(M_H>\unit[150]{GeV}\), which is clearly excluded. We discuss also the prospects for differentiating different boundary conditions imposed for \(\lambda(M_{pl})\) at the LHC. A striking example is \(M_H = \unit[127\pm 5]{GeV}\) corresponding to \(\lambda(M_{pl})=0\), which would imply that the quartic Higgs coupling at the electroweak scale is entirely radiatively generated. 
\end{abstract}
\maketitle

\section{Introduction \label{sec:outline}}
The Standard Model (SM) of particle physics is very successful and it has withstood all precision tests over almost 40 years. All SM particles have been discovered\footnote{In addition neutrinos were found to be massive, which requires in its simplest form only the addition of three right handed fermionic singlets.}, except for the ingredient connected to electroweak Symmetry Breaking (EWSB), namely the Higgs boson. The Large Hadron Collider (LHC) was built to test EWSB and to find or rule out the Higgs boson. In addition, the LHC aims at detecting new physics which is suggested to exist in the TeV range in order to solve the so-called hierarchy problem. However, the ATLAS and CMS detectors at LHC have so far not found any sign of new physics and the remaining Higgs mass range has shrunk considerably \cite{ATLAS-CONF-2011-157}. This suggests to think about scenarios where nothing but the SM is seen. 

The essence of the hierarchy problem \cite{Weinberg:1975gm,*Weinberg:1979bn,*Gildener:1976ai,*Susskind:1978ms,*tHooft:1980xb} is the fact that quantum corrections generically destroy the separation of two scales of scalar Quantum Field Theories (QFT). It is thus not possible to understand how the electroweak scale could be many orders of magnitude smaller than the scale of an embedding QFT. Conventional solutions of the hierarchy problem stay within QFT. One solution is to postulate a new symmetry (Supersymmetry) which cancels the problematic quadratic divergences. Alternatively the scalar sector may be considered effective (composite) such that form factors remove large quadratic divergences. Another idea is that the Higgs particle could be a pseudo-Goldstone-boson such that its mass is naturally somewhat lower than a scale where richer new physics exists. However, none of these ideas has so far shown up in experiments. 

This prompts us to consider new ideas. An important observation is the fact that the SM can anyway not be valid up to an arbitrary high energy scale due to triviality \cite{Lindner:1985uk} and since gravity must affect elementary particle physics latest at the Planck scale, \(M_{pl}=\unit[1.2\times10^{19}]{GeV}\). Note that this introduces a conceptual asymmetry, as gravity is known to be non-renormalizable, i.e. it {\em cannot} be a QFT in the usual sense and requires fundamentally new ingredients. This looks bad from the perspective of renormalizable gauge theories, also since we can so far only guess which concepts might be at work. However, this may also be good in two ways: First, renormalizable QFTs do not allow to calculate absolute masses and absolute couplings. Any embedding of the SM into some other renormalizable QFT (like GUTs) would therefore only shift the problem to a new theory which is also unable to determine the absolute values of these quantities. In other words: The problems of  gravity may be a sign of physics based on new concepts which may ultimately allow to determine absolute masses, mixings and couplings. However, there is no need for the SM to be directly embedded into gravity and various layers of conventional gauge theories (LR, PS, GUT, ...) could be in-between. The second reason why an embedding involving new concepts beyond renormalizable QFTs might be good is that this asymmetry might offer new solutions to the hierarchy problem. In other words: The conventional approach towards interpreting quadratic divergences to the Higgs mass correction by simply substituting \(\Lambda^2\rightarrow M_{pl}^2\) might not be correct if Planck scale physics is based on new physical concepts different from conventional QFT. This view is also shared by Meissner and Nicolai \cite{Meissner:2007xv}. The point is that the unknown new concepts may allow for mechanisms which stabilize a low-lying effective QFT from the perspective of the Planck scale. From the perspective of the low-lying effective QFT this may then appear to be a hierarchy problem if one tries to embed into a renormalizable QFT instead of the theory which is based on the new, extended concepts.
\setlength{\unitlength}{5cm}
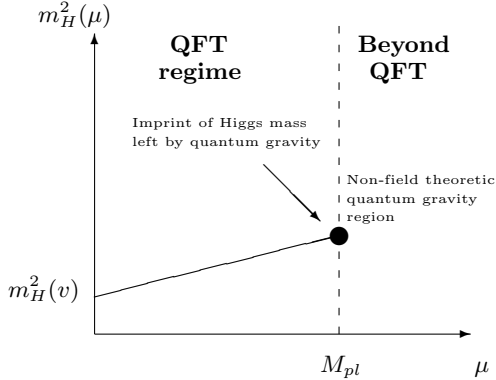
\begin{figure}
\begin{center}
\begin{picture}(1,1)
\put(0,0.1){\vector(0,1){0.8}}
\put(0,0.1){\vector(1,0){1}}
\put(1.01,0){$\mu$}
\put(-0.15,0.93){$m_H^2(\mu)$}
\put(0,0.2){\line(4,1){0.65}}
\put(-0.23,0.2){$m_H^2(v)$}
\put(0.6,0){$M_{pl}$}
\multiput(0.65,0.1)(0,0.05){17}
{\line(0,1){0.02}}
\put(0.67,0.5){\tiny Non-field theoretic}
\put(0.67,0.45){\tiny quantum gravity}
\put(0.67,0.4){\tiny region}
\put(0.45,0.55){\vector(1,-1){0.15}}
\put(0.65,0.36){\circle*{0.05}}
\put(0.1,0.65){\tiny Imprint of Higgs mass}
\put(0.1,0.6){\tiny left by quantum gravity}
\put(0.2,0.85){\textbf{QFT}}
\put(0.17,0.78){\textbf{regime}}
\put(0.7,0.85){\textbf{Beyond}}
\put(0.73,0.78){\textbf{QFT}}
\end{picture}
\end{center}
\label{fig:remnant}
\caption{The SM Higgs mass could be determined and fixed by unknown physics connected to quantum gravity, which should be based on new concepts other than conventional QFT. The running of the Higgs mass from Planck scale down to electroweak scale is fully dictated by the SM as a QFT. }
\end{figure}
\setlength{\unitlength}{1pt}

In condensed matter physics for instance, the energy of a superconductor in Ginzburg-Landau theory is described by:
\begin{align}
E&\approx \alpha |\phi|^2 +\beta |\phi|^4+\ldots 
\end{align}
where \(\alpha\) and \(\beta\) are phenomenological parameters. These parameters have to be determined by experiment itself in the Ginzburg-Landau theory framework, but they can be calculated from the microscopic theory of superconductivity, namely the BCS theory. In this sense the microscopic theory fixes the parameter or the boundary condition of the low energy effective theory. The parameters in low energy Ginzburg-Landau theory ``know'' the boundary condition set by the underlying BCS theory, but many dynamical details of the BCS theory are lost in the Ginzburg-Landau effective theory, even though BCS theory does not provide a mechanism to explain hierarchies. 

The above considerations prompt us to speculate that the SM might be valid up to the Planck scale, where it is embedded directly into Planck scale physics without any intermediate energy scale. The new concepts behind the Planck scale physics might then offer a solution to the hierarchy problem which is no longer visible when one looks at the SM only. In analogy to the  superconductivity example the only way how the SM would ``know'' about such an embedding could be special boundary conditions similar to compositeness conditions or auxiliary field conditions in theories where redundant degrees of freedom are eliminated in embeddings. This scenario is depicted in Fig.~(\ref{fig:remnant}).

Following the logic outlined above it would be essential to have only the weak and the Planck scale and nothing in between, since otherwise it would require to solve the hierarchy problem within QFT. We therefore forbid any kind of new intermediate energy scale between weak and Planck scale in order to avoid the large hierarchy between Higgs and heavy intermediate particle's mass. This view is also shared by the spirit of \(\nu\mathrm{MSM}\), proposed by Shaposhnikov \cite{Shaposhnikov:2007nj}. Even without any intermediate scale one might still ask why there should not be any radiative corrections of Planck scale size to the Higgs mass parameter. We do not have an argument here but we point the interested reader to the works of Bardeen \cite{Bardeen:1995kv} that have argued about the spurious nature of the quadratic correction. 

In the logic of the above arguments we assume therefore in this letter simply that the SM is valid up to the Planck scale, and that quantum gravity leaves certain boundary conditions for the Higgs quartic coupling \(\lambda\).

\section{Boundary conditions on \texorpdfstring{$\lambda$}{lambda} at the Planck scale\label{sec:boundary}}
It is well known that the SM cannot be extrapolated to arbitrarily high energies. A first constraint comes from the fact that at very high energies far above the Planck scale the \(U(1)_Y\) gauge coupling diverges. A more important piece of theoretical input comes from the SM Higgs sector: for small Higgs masses of approximately less than \(\unit[127]{GeV}\), the contributions from top loops drive the Higgs self-coupling towards negative values before the Planck scale and thus make the Higgs potential unstable \cite{Lindner:1988ww,Sher:1988mj}. For Higgs masses larger than approximately \(\unit[170]{GeV}\), the contribution from the Higgs self-coupling drives itself towards a Landau pole before the Planck scale, this is the so-called \emph{triviality bound} \cite{Lindner:1985uk}.

Suppose that the LHC only detects the SM Higgs boson with a mass in the range between \(\unit[127-150]{GeV}\) and nothing else. The triviality and vacuum stability bounds then imply that the SM can be effectively valid up to the Planck scale. The SM parameters could then be directly determined by Planck scale physics and one might ask the question if there is a way to gain information on this type of physics from measurements performed at LHC. Motivated by an asymptotic safety scenario of gravity, Shaposhnikov and Wetterich \cite{Shaposhnikov:2009pv} have, for example, proposed that both the Higgs self-interaction $\lambda$ and its beta function $\beta_{\lambda}$ should simultaneously vanish at the Planck scale, from which they derive the prediction of $m_H=126$ GeV. At first sight it seems remarkable that both conditions can be fulfilled at the same time and this prompted us to look at such type of boundary conditions in more detail.

We discuss therefore the following boundary conditions which we imagine to be imposed on the SM in the spirit of this paper by some version of quantum gravity\footnote{Boundary conditions of this type have also been discussed in the context of anthropic considerations in the multiverse \cite{Hall:2009nd}.}:
\begin{itemize}
\renewcommand{\labelitemi}{$\bullet$}
\item Vacuum stability\\ \(\lambda(M_{pl})=0\) \cite{Froggatt:1995rt,Shaposhnikov:2009pv,Casas:1994qy,Casas:1996aq,Sher:1988mj,Lindner:1988ww,Isidori:2007vm}.
\item vanishing of the beta function of \(\lambda\)\\ \(\beta_{\lambda}(M_{pl})=0\) \cite{Froggatt:1995rt,Shaposhnikov:2009pv}.
\item the Veltman condition  \\ $\mathrm{Str}\mathcal{M}^2=0$ \cite{Chaichian:1995ef,Veltman:1980mj,Decker:1979nk}, \\ which states that the quadratic divergent part of the one-loop radiative correction to the Higgs bare mass parameter $m^2$ should vanish\footnote{Note that the notation of \(\mathrm{Str}\mathcal{M}^2\) for the Veltman condition includes only the correction to \(m\) by the running coupling which is not the direct matching of respective pole masses. Note also, that we have only included the top quark Yukawa coupling \(\lambda_t\) and omitted the other Yukawa couplings, as they do not contribute significantly to the Higgs mass running compared to the contributions from  \(\lambda\), \(\lambda_t\), and \(SU(2)_L\times U(1)_Y\) gauge couplings \(g_2\) and \(g_1\). It is known that the Veltman condition is scheme dependent as the quadratic divergence from different particles is not necessarily the same. However if we assume a common cut-off for all the particle contributions, which may appear appropriate for our scenario, we will obtain a range of Higgs masses which is still not excluded by the experiments.}:
\begin{align}
\delta m^2 &=\frac{\Lambda^2}{32\pi^2 v^2} \mathrm{Str}\mathcal{M}^2
\label{eq:strm} \displaybreak[0] \\
&=\frac{1}{32\pi^2}\left(\frac{9}{4}g_2^2+\frac{3}{4}g_1^2+6\lambda-6\lambda_t^2  \right)\Lambda^2.
\label{eq:veltman}
\end{align}
\item vanishing anomalous dimension of the Higgs mass parameter\\ \(\gamma_m(M_{pl})=0,\, m(M_{pl})\neq0\).
\end{itemize}

\begin{figure*}
\centering
\includegraphics[width=.8\textwidth]{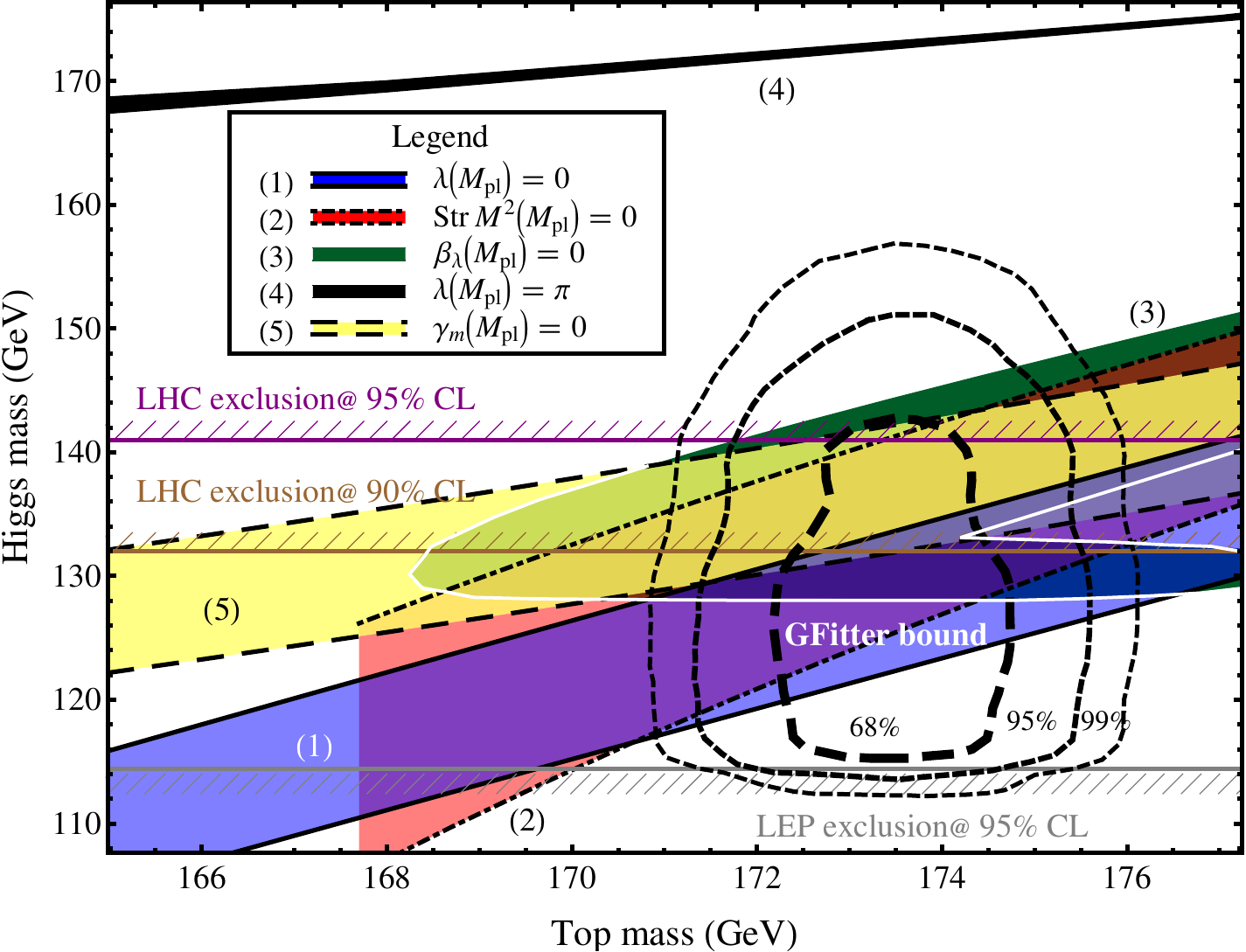}
\caption{Higgs and top (pole) mass determinations for different boundary conditions at the Planck scale. The coloured bands correspond to the conditions discussed in the text and which are also labelled in the insert. The middle of each band is the  best value, while the width of the band is a ``RGE error band'' inferred from assuming that all omitted higher orders in the beta functions beyond two loops are limited by the difference between the one and two loop results. Note that the Veltman condition is truncated at the point where its Higgs mass prediction violates the vacuum stability bound (both at two-loops). The gray-hatched line at the bottom is the lower direct Higgs mass bound from LEP.  Similarly the purple (brown) lines indicate the LHC Higgs searches at 95\% (90\%) CL from the 2010 data. The black dashed lines show the electroweak precision fit from GFitter \cite{Baak:2011ze,Bardin:1999yd,*Arbuzov:2005ma} for 68\%, 95\% and 99\% confidence intervals (which include limits from radiative corrections and also the direct searches). }
\label{fig:pd1} 
\end{figure*}

As we aim to determine the Higgs mass due to different boundary conditions for \(\lambda\) imposed at the Planck scale, we use renormalization group equations (RGEs) to evolve the couplings. The relevant one- and two-loop beta functions required for solving the RGE are listed in App.~(\ref{sec:a1}). As the uncertainty in the top mass is the dominant source of uncertainty for the resulting Higgs mass prediction we treat the top mass as a free parameter (within a certain range) and show the dependence on the top mass explicitly. For each top mass value we determine the corresponding Higgs mass due to a given boundary condition on \(\lambda(M_{pl})\). For that we need to convert the top pole mass to its corresponding \(\overline{\mathrm{MS}}\) Yukawa coupling:
\begin{equation}
\lambda_t(M_t)=\frac{\sqrt{2}M_t}{v}\left(1+\delta_t(M_t) \right),
\label{eq:toppole}
\end{equation}
where \(\delta_t(M_t)\) is the matching correction for top mass. The list of matching conditions used for \(\delta_t\) is given in App.~(\ref{sec:a2}). The matching scale is chosen to be \(\mu=M_t\), which is a suitable choice for a low Higgs mass range \cite{Hambye:1996wb}. We consider also the threshold effects in the beta functions: The known gauge couplings \(g_i(M_Z)\) run to the scale \(\mu=M_t\) without including the top loop contribution, and then the value of  \(g_i(\mu=M_t)\) will be used in subsequent RGE, where we compute all the coupled differential equation of \(g_i\), \(\lambda\) and \(\lambda_t\) with boundary condition imposed. With suitable boundary conditions imposed for \(\lambda\) at the Planck scale, we can extract the Higgs mass at \(\mu=M_t\) after solving the RGE. The \(\overline{\mathrm{MS}}\) Higgs quartic coupling is then matched to the pole mass with:
\begin{align}
\lambda(M_t)=\frac{M_H^2}{2v^2}(1+\delta_H(M_t)),
\label{eq:higgspole}
\end{align}
where the Higgs matching \(\delta_H\) is given at App.~(\ref{sec:a2}). Repeating the procedure for different values of the top pole mass, we obtain the Higgs mass determinations plotted in Fig.~(\ref{fig:pd1}).

Next we discuss some details regarding the boundary condition imposed for \(\lambda\) at the Planck scale, starting with the vacuum stability bound. To obtain the vacuum stability bound we need to consider two cases in solving the coupled differential equations \cite{Casas:1994qy,Casas:1996aq}:
\begin{enumerate}
\item We first impose the boundary conditions at tree-level, i.e. \(\lambda(M_{pl})=0\) and apply the one-loop beta functions and anomalous dimension equations in solving the RGEs numerically. 
\item Two-loop beta functions and anomalous dimension for \(m\) are considered in our RGE and the effective potential is considered in one-loop approximation. The condition that we would need to impose is:
\begin{align}
\lambda(M_{pl})&=\frac{1}{32\pi^2}\left(\frac{3}{8}(g_1^2(M_{pl})+g_2^2(M_{pl}))^2\left[\frac{1}{3}\right.\right. \nonumber \\ 
&\quad\left.-\log\frac{g_1^2(M_{pl})+g_2^2(M_{pl})}{4}\right]+6\lambda_t^4(M_{pl})  \nonumber \\ &\quad\left[\log\frac{\lambda_t^2(M_{pl})}{2}-1\right]+\frac{3}{4}g_2^4(M_{pl})\left[\frac{1}{3}\right. \nonumber \\
&\quad\left.\left.-\log\frac{g_2^2(M_{pl})}{4}\right]\right)
\end{align}
\end{enumerate}
Effectively we want to be consistent with the vacuum stability bound obtained from the effective potential. With this approach we can estimate the uncertainty from the difference of Higgs masses obtained via different cases above, which is effectively due to the omission of higher-order contributions to the beta functions and correction to the effective potential. 

For the case of the second boundary condition, we only impose the tree level Veltman condition given in Eq.~\eqref{eq:strm}, as higher order loops always come with \(\log(M_{pl}/\mu)\) \cite{Einhorn:1992um}, which will be cancelled if the running couplings are evaluated at Planck scale \cite{Kolda:2000wi}. This is true, however, only if the complete beta functions for all the couplings are used to resum the complete order of logarithms.

The gray hatched line in Fig.~(\ref{fig:pd1}) depicts the lower direct Higgs mass search bound from LEP  \cite{Barate:2003sz}. The coloured hatched lines give the combined exclusion limits from ATLAS and CMS  of \(\unit[141-476]{GeV}\) at 95\% confidence level (CL) and \(\unit[132-476]{GeV}\) at 90\% CL \cite{ATLAS-CONF-2011-157}.  The experimentally favoured parameter range in the $m_t$--$m_h$ plane taking into account direct Higgs searches and and electroweak precision measurements are indicated by the GFitter region \cite{Baak:2011ze,Bardin:1999yd,*Arbuzov:2005ma} in the plot. We show the dependence on the top mass outside of the range \(\unit[172.3-174.1]{GeV}\), corresponding to the best world average value of top quark mass \(\unit[173.2\pm0.9]{GeV}\) \cite{Lancaster:2011wr}, as methods that determine the top mass directly from the $t\overline{t}$ cross-section favour a smaller value of \(\unit[168.9^{+3.5}_{-3.4}]{GeV}\) \cite{Langenfeld:2009wd}.

We observe that most of the Higgs masses given by different conditions tend to overlap in the vicinity of the best determined value of the top mass. The triviality bound, represented by the approximate condition \(\lambda(M_{pl})=\pi\), yields a range of Higgs masses which is already excluded at 95\% CL by the Tevatron and LHC. The Higgs masses generated by the other  conditions however are still allowed and not excluded yet by the experiments. The Veltman condition is truncated at the point where its Higgs mass calculated with two-loop beta functions starts to cross the vacuum stability bound obtained by two-loop RGEs. This is done in order to show the exact crossing point of these two conditions from two-loop RGEs. 

Throughout this work, we define a measure of the uncertainties involved in the calculation as the difference between using one and two-loop beta functions for all the relevant SM couplings in the determination of the Higgs pole mass. To be more precise, we define a ``RGE error band'' as the difference in determining the Higgs mass for a boundary condition of \(\lambda(M_{pl})\) with one and two-loop beta functions while the matching conditions remain the same for both cases. Possible errors due to the matching conditions will be discussed below.  We caution the reader that this procedure probably overestimates the error stemming from neglecting higher order contributions to the beta functions. While there is no universally accepted way of estimating the theoretical uncertainties\footnote{See \cite{Cacciari:2011zr} for a recent attempt at rigorously characterizing a perturbative theoretical uncertainty.} , there are other approaches to define the theoretical error used in the literature. E.g. in Ref.~ \cite{EliasMiro:2011aa}, the authors define the theoretical error by the scale dependence of the matching condition $\lambda(M_t)$ and $\lambda_t(M_t)$ while neglecting the effect of higher order RGEs and arrive at an 3 GeV uncertainty. In Ref. \cite{Bezrukov:2009fr}, the authors estimate the theoretical uncertainty by comparing the situation where matching has been performed at $\mu=M_t$ to the case $\mu=M_Z$ and get an uncertainty of $2.2$ GeV. We comment on possibilities  to reduce the theoretical uncertainty at the end of the paper. If one rather believes in this method of error estimation, the error curves shink accordingly. In our plot, the aforementioned ``RGE error band'' is represented by the bandwidth of each curve, with its center representing the Higgs mass obtained from two-loop RGE running. The upper edges of the bandwidths consist of the Higgs masses obtained from one-loop RGEs.

We consider also the uncertainty on the curves due to the error of strong coupling constant \(\alpha_s=0.1184(7)\) \cite{Nakamura:2010zzi} and we obtain \(\pm \unit[1]{GeV}\) uncertainty to the Higgs mass, which is negligible when quadratically added to the bandwidth in Fig.(\ref{fig:pd1}). Due to the relatively large ``RGE error band'', the error propagation from the strong coupling constant can be safely ignored. The theoretical error on the Higgs mass due to the matching uncertainty \cite{Hambye:1996wb,Hempfling:1994ar} between top Yukawa \(\overline{\mathrm{MS}}\) coupling and top pole mass is also considered. Comparing our vacuum stability band obtained with Casas et al.\ \cite{Casas:1994qy,Casas:1996aq}, a discrepancy of around \(\pm\unit[7]{GeV}\) for the Higgs mass value obtained via two-loop RGE is observed. This mismatch can be explained by the omission of two-loop QCD matching condition by the authors of Refs.~\cite{Casas:1994qy,Casas:1996aq}, as they only considered one-loop QCD, electroweak and QED contribution in the top mass matching condition. Since we would like to consider only the uncertainties due to the number of loops of the beta function used but not the errors caused by omission of better matching precision, we include the QCD matching between top Yukawa \(\overline{\mathrm{MS}}\) coupling and top pole mass up to three-loop. The resulting Higgs mass determined by the vacuum stability with two-loop RGE agrees with Ellis et al.\ \cite{Ellis:2009tp}. The \(\alpha\alpha_s\) correction \cite{Jegerlehner:2003sp} is neglected in our analysis as it only gives a small contribution. The Higgs pole mass is matched with \(\lambda\) at top pole mass scale, i.e.\ the renormalization scale of \(\lambda\) is set to be at \(\mu=M_t\). Since the higher order matching conditions for \(\lambda\) has not been calculated in any literature, only the expression of \(\delta_H\) given in App.~(\ref{sec:a2}) will be used as our matching. If the matching of \(\lambda\) to the Higgs pole mass is only performed at tree-level, the resulting discrepancy of the Higgs pole mass with respect to the one-loop matching result is found to be less than \(\unit[1]{GeV}\). Therefore, we can safely assume that higher order matching condition for \(\lambda\) will not yield a larger uncertainty.

The error estimation for the boundary condition \(\beta_{\lambda}(M_{pl})=0\) is not the same as for the other conditions. A careful treatment of the Higgs mass extraction has to be implemented in this case. The one-loop beta function of \(\lambda\) is a quadratic equation of \(\lambda\), and for a given top mass we obtain two positive solutions of the boundary condition \(\beta_{\lambda}(M_{pl})=0\) at the Planck scale,  and thus obtain two equally valid low-energy Higgs mass determinations. In Fig.(\ref{fig:pd1}) these two branches of solutions generate a hook-shaped trajectory in the \(M_H-M_t\) plane. The hook ends where \(\lambda\) starts to take negative value. Due to the mismatch of the end of the trajectory when either one-loop or two-loop beta function is applied, a larger ``RGE error band'' has to be taken into account, where we generate error bars that cover the distance of the mismatch and plot a band to cover all the error bars. Besides the mismatch mentioned above, there exists also another source of error, namely number of loops of the beta functions implemented in the condition \(\beta_{\lambda}(M_{pl})=0\). In principle one should apply the full beta function as the boundary condition, but in practice this is impossible and therefore we have to check the possible uncertainty which arises due to the number of loops in \(\beta_{\lambda}\) used in the boundary condition. The errors lie, however, within the band. A similar uncertainty due to number of loops used as boundary condition also appears in the \(\gamma_m(M_{pl})=0\) condition and its uncertainty is larger in comparison to the \(\beta_{\lambda}(M_{pl})=0\) condition.  

Another possible source of uncertainties on the determination of the Higgs mass from the Planck scale boundary conditions comes from the value of the Planck scale. The difference of the Higgs masses obtained from a certain boundary condition imposed at a different value of the Planck scale, e.g.\ \(\mu=M_{pl}/\sqrt{8\pi}\), are negligible for most of the boundary conditions. However the discrepancies are larger for the lower branch of the \(\beta_{\lambda}=0\) condition. We will not discuss further these uncertainties.

Throughout this work we assume that neutrinos do not play a significant role in our Higgs mass prediction. We would like to caution the reader that neutrinos are indeed massive and could couple with the \(\mathcal{O}(1)\) Yukawa coupling to the Higgs, e.g. in the canonical type I see-saw mechanism. For a neutrino mass of \(m_{\nu}\approx \mathcal{O}(\unit[1]{eV})\) and a see-saw scale of approximately \(\unit[10^{15}]{GeV}\), the Yukawa coupling of neutrino would be around \(\mathcal{O}(1)\). This large coupling between Higgs boson and neutrino could alter the prediction on Higgs mass by modifying the beta function of Higgs quartic coupling, as pointed out by Ref.~\cite{Casas:1999cd}. However the interesting case, i.e. SM plus extension of neutrino sector with \(\mathcal{O}(1)\) Yukawa coupling can be valid up to Planck scale, has been excluded by recent Tevatron and LHC searches \cite{ATLAS-CONF-2011-157}. We therefore implicitly assume a scenario where the neutrino Yukawa couplings are not significantly larger than for example the bottom Yukawa coupling.

\section{Is a low Higgs mass favoured? \label{sec:probability}}

\begin{figure}
\includegraphics[scale=0.9]{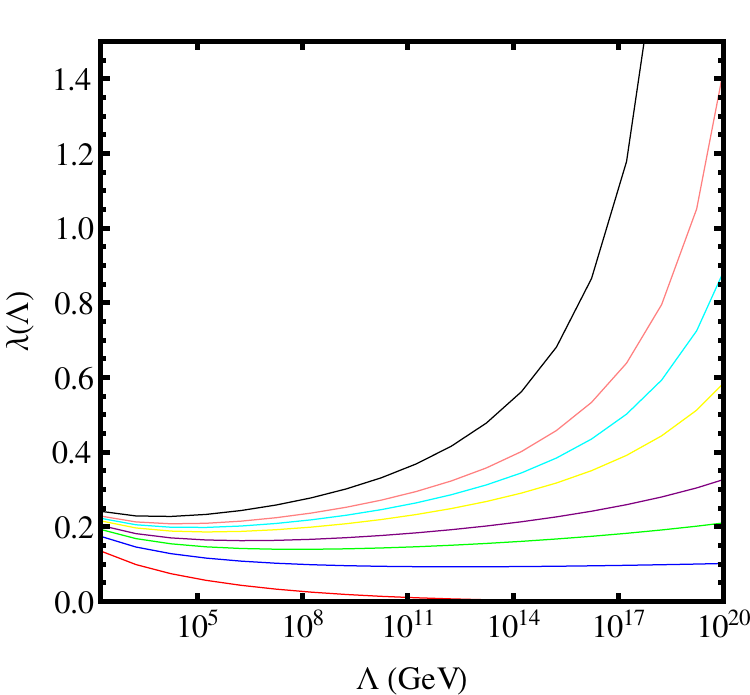}
\caption{Running of \(\lambda\) from the Planck scale to the Fermi scale. Different values of \(\lambda(M_{pl})\) and it can be seen that a large parameter space of \(\lambda(M_{pl})\) tends to induce \(\lambda(M_t=\unit[173]{GeV})\gtrsim\, 0.186\) which is equivalent to Higgs mass greater than \(\unit[150]{GeV}\).}
\label{fig:run} 
\end{figure}

All the generic boundary conditions which we discussed prefer a low Higgs mass \(\unit[127]{GeV} \lesssim M_H\lesssim \unit[145]{GeV}\) which fits amazingly good to the existing experimental direct lower and upper bounds from LEP, Tevatron and LHC. These values fit furthermore very well to those preferred by electroweak precision measurement. One could ask therefore if this is by chance or if it has a specific reason. Let us therefore first look at random values of \(\lambda(M_{pl})\) pretending in this way that every value could be realized by the new physics at the Planck scale. We assume therefore for a moment, that all values of \(\lambda(M_{pl})\) in the range of \(\lambda\in[0,\pi]\) have equal probability\footnote{In principle one could consider all values up to infinity since there seems to be no a priori reason to limit \(\lambda(M_{pl})\). Extending the upper end is however only strengthening the arguments, since this would favour even more a higher Higgs mass. Note that this is connected to the triviality bound and the corresponding focussing of RGE trajectories towards low scales. See Fig.(\ref{fig:run}) for this effect.}. We show therefore first in Fig.(\ref{fig:run}) the running of  \(\lambda\) from the Planck scale to the weak scale for some values in the interval  \(\lambda\in[0,\pi]\). Note that most values of \(\lambda(M_{pl})\) lead to \(\lambda\) at the Fermi scale which is greater than 0.2, which corresponds to \(M_H>\unit[150]{GeV}\). 

\begin{figure}
\includegraphics[scale=0.6]{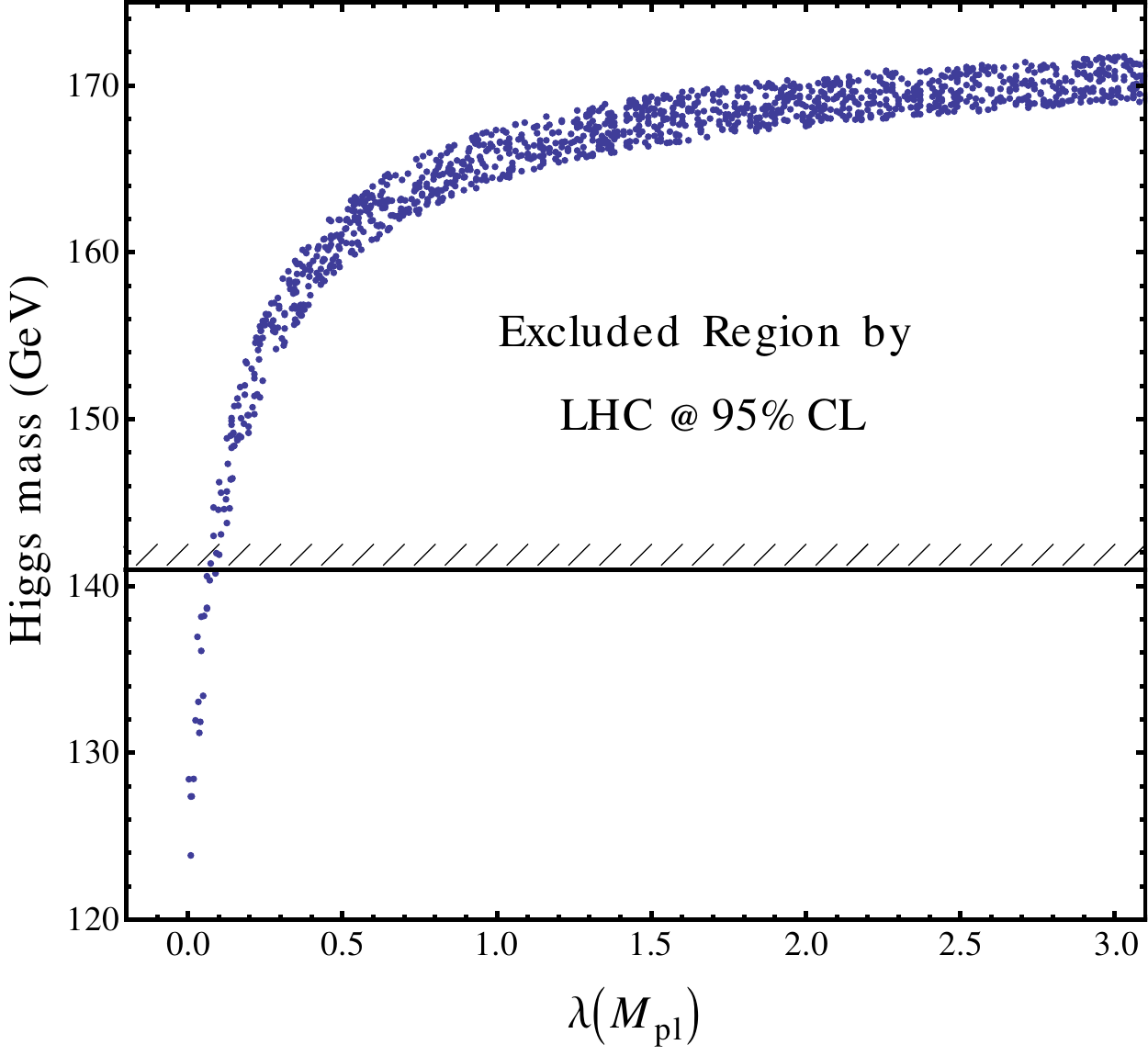}
\caption{Scatter plot of Higgs mass at the Fermi scale determined by random \(\lambda\) at the Planck scale with random top mass constrained to the interval \(\left[170,175\right]\) GeV.}
\label{fig:histo} 
\end{figure}

To analyse the effect further we randomly generate 600 values for \(\lambda(M_{pl})\) and the top pole mass ranging from \(\unit[170-175]{GeV}\) and put the resulting Higgs masses into the scatter plot Fig.(\ref{fig:histo}). Note that most values of  \(\lambda\) at the Planck scale lead to a Higgs mass between \(\unit[160]{GeV}\) and \(\unit[175]{GeV}\). For the chosen interval \(\lambda(M_{pl})\in[0,\pi]\) we find \(\approx90\%\) in this range which has been excluded by the Tevatron and by the CMS and ATLAS experiments. Note that only less than 5\% of the generated Higgs masses are allowed by experiments. 

Looking at Fig.(\ref{fig:histo}) one immediately notices that \(\lambda(M_{pl}) = 0\) which corresponds to \(M_H = \unit[127]{GeV}\) is a special condition which fits very well to the experimental findings. This is also the vacuum stability bound and it corresponds therefore to the lightest Higgs mass in the SM when it is valid up to the Planck scale. A lower Higgs mass would require some extra new physics at lower scales, which would rule out the logic of our paper. Including errors this leads to a lower bound \(M_H \gtrsim \unit[122]{GeV}\) for our scenario. 

Fig.(\ref{fig:histo}) also shows that it is non-trivial that all the boundary conditions which we discuss in this paper lead to viable Higgs masses. However, there is a systematic understanding why all our conditions work. The point is that all our conditions such as \(\beta_{\lambda}(M_{pl})=0\) or \(\mathrm{Str}\mathcal{M}^2(M_{pl})=0\) are conditions connected to quantum loops for the Higgs mass parameter \(m\) and the quartic coupling \(\lambda\) and we can therefore always write \(\lambda\) as a function of the gauge couplings and the top Yukawa coupling, i.e. \(\lambda=f(g_i,\lambda_t)\). This leads generically to smaller values than a random choice, since loop factors, i.e. factors of \(1/16\pi^2\) are present and since top mass loops (fermion loop) carry a minus sign, which leads to cancellations, pushing the value of \(\lambda\) even smaller. One can therefore understand why the cases which we discussed all systematically lead to viable Higgs mass predictions. In that sense none of them is special and there may exits more interesting boundary conditions which also lead to viable Higgs masses. This also implies that a Higgs mass in the currently most favoured range does not clearly select any model or scenario which leads to a boundary condition that has such loop and top mass suppression factors. However, vice versa one might argue that current data point to boundary conditions which must involve such suppression factors, which is an interesting observation. It is also intriguing to ask in this context why the top mass is much heavier than other quarks such that it compensates other loop contributions which would drive the Higgs boson heavier. This appears to be an interesting ``conspiracy'' in favour of the logic of this paper which works if the boundary condition is imposed at a high scale like the Planck scale. All these arguments break of course down if LHC or any future experiment detects any sign of new physics that couples directly to the Higgs boson. Even an indirect coupling, i.e. radiative correction to \(\lambda\) at loop level is severe enough to alter the running of \(\lambda\) drastically. So far there is no compelling evidence for additional physics beyond the SM, whereas the SM Higgs search seems to indicate some excesses of Higgs like events in the range of \(\unit[130]{GeV}\) to \(\unit[140]{GeV}\), albeit at only about \(2\sigma\) \cite{ATLAS-CONF-2011-157}. 

\section{Discussion and conclusions \label{sec:merge}}

The LHC has an excellent chance to find the SM Higgs boson and we emphasize in this paper that the left-over values lie in a range which is well motivated by various Planck scale boundary conditions. We argued that this Higgs mass range is special and that it might be related to embeddings of the SM as a QFT into some form of quantum gravity, which is based on concepts {\em beyond} QFT. The SM would still be an effective field theory which is valid up to the Planck scale, but the asymmetry in the concepts might allow to understand the famous hierarchy problem from the perspective of the new concepts at the Planck scale, while it would only appear unnatural from the low energy point of view. In other words: The large hierarchy between the Planck and electroweak scale might only be a problem as long as we look at it from the SM perspective. An important point is then that such a scenario makes only sense if there are no intermediate scales, since this would require a QFT solution of the hierarchy between the electroweak and intermediate scales. 

The proposed scenario requires that the Higgs coupling can be evolved to the Planck scale, which implies strict lower and upper bounds on the coupling. The upper bound is the so-called triviality bound which is approximately \(\unit[170]{GeV}\) and it is interesting to note that the Higgs mass is below this value, as otherwise the couplings could not be evolved up to the Planck scale. The lower bound is the so-called vacuum stability bound, i.e. the condition \(\lambda(M_{pl})=0\). We carefully evaluated its value and error with the latest data at two loops and provided an error estimate. For \(M_t=\unit[173.2]{GeV}\) we find  \(m_H = \unit[127\pm 5]{GeV}\) which implies that the Higgs mass must be heavier than \(\unit[122]{GeV}\). 
It is important to note that the vacuum stability bound, or equivalently the condition \(\lambda(M_{pl})=0\), is very special. It implies that the Higgs self-interaction at the electroweak scale is entirely generated by radiative corrections of the RGE evolution from a vanishing coupling at the Planck scale. The Higgs mass would therefore be connected to the gauge and Yukawa couplings which enter into the RGEs. 

\begin{figure}
\includegraphics[scale=0.7]{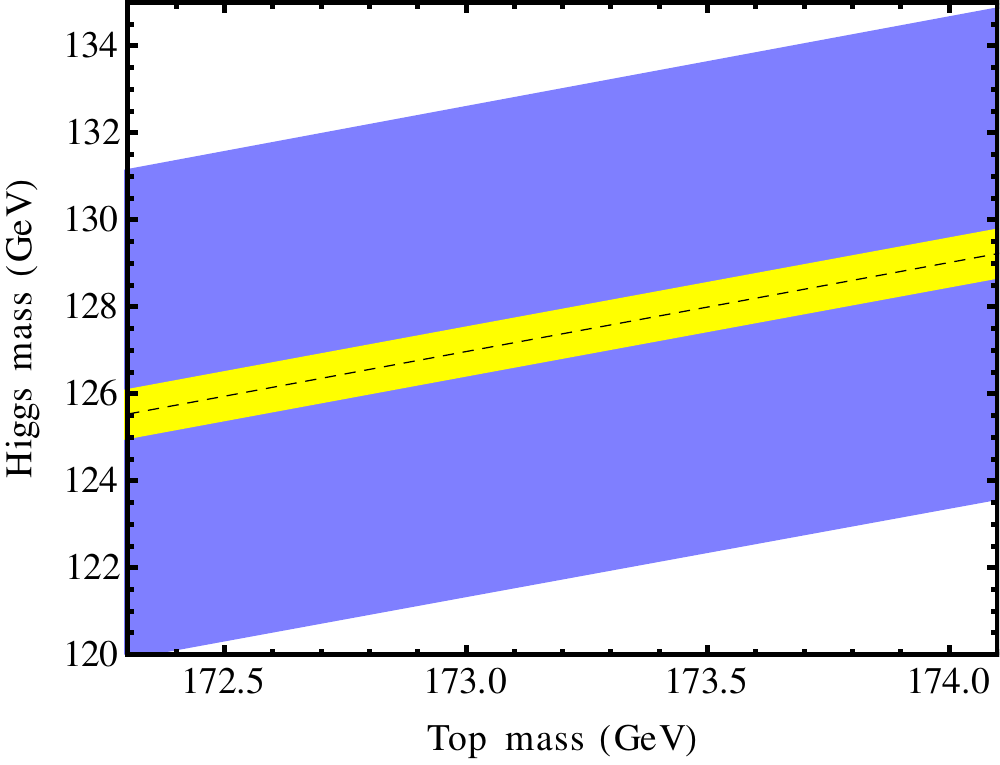}
\caption{A blow up of the vacuum stability bound in the interesting Higgs and top mass region. The blue line in the center represents the vacuum stability bound obtained via two-loop beta functions, which has been thoroughly discussed in main text. The yellow band represents the uncertainties of the Higgs mass obtained via two-loop RGEs due to \(\alpha_s\) uncertainties. The outer blue band is identical to the blue band in Fig.~(\ref{fig:pd1}) and it represents the full ``RGE error band'' estimated from difference between one- and two-loop RGEs. With the best world average top pole mass \(\unit[173.2]{GeV}\) the inferred Higgs mass from the vacuum stability condition \(\lambda(M_{pl})=0\) is \(\unit[127 \pm 5]{GeV}\).}
\label{fig:vacuumerror}
\end{figure}
Several comments should be carefully considered in this context:
\begin{enumerate}
\item The Higgs central value \(m_H = \unit[127]{GeV}\) is obtained via two-loop beta function running from the vacuum stability condition at the Planck scale to the weak scale regime. Fig.~(\ref{fig:vacuumerror}) shows the uncertainties due to the omission of higher orders to be \(\pm\unit[5]{GeV}\). This appears a reasonable way to arrive at a conservative error estimation for the Higgs mass due to the lack of higher order RGEs, but it should not be over interpreted. This implies that the exact lower bound for the Higgs mass is limited by this conservative estimation. 
\item Precision top mass analysis is required to determine the exact value of the Higgs mass predicted via vacuum stability. The reason why we want to stress on this specific result is that to date, there is no general consensus on what type of top mass is actually measured via kinematic reconstruction \cite{hoang:lec}. At the Tevatron, the main method used for the top mass extraction actually ``measures'' the Pythia mass, which is a Monte-Carlo simulated template mass. Strictly speaking the top pole mass is not a well defined quantity, as the top quark does not exist as free parton. The top mass that the Tevatron has measured is based on the final state of the decay products. On the other hand the running \(\overline{\mathrm{MS}}\) top mass can be extracted directly from the total cross section in the top pair production. In this sense, one can obtain a complementary information of the top mass from the production phase. By converting the \(\overline{\mathrm{MS}}\) mass to the pole mass via matching conditions, the top pole mass value \(\unit[168.9^{+3.5}_{-3.4}]{GeV}\) extracted with this method by Langenfeld et al.\ \cite{Langenfeld:2009wd} is found to be lower than the world best average value. 
However, this way of extracting the top mass suffers from larger numerical uncertainties. As we can be seen from Figs.~(\ref{fig:pd1}) and (\ref{fig:vacuumerror}), a change of the top mass by \(\unit[2]{GeV}\) changes the Higgs mass prediction by \(\unit[6]{GeV}\).
%
%
\item The electroweak vacuum might in principle be metastable. However, most of the Higgs mass region for metastability has already been ruled out by LEP, although not entirely excluded. The finite temperature metastability region, however, with the local SM assumed to be stable against thermal fluctuations up to Planck scale temperatures, allows the entire region from the LEP bound to the vacuum stability. 
Hence should LHC discover the SM Higgs boson with its mass lower than the one predicted by two-loop RGE vacuum stability bound, there is a possibility that the SM electroweak vacuum is not the stable one (see Refs.~\cite{Ellis:2009tp,Espinosa:2007qp,Isidori:2001bm,Isidori:2007vm,Froggatt:2001pa,EliasMiro:2011aa} for more details). However, we would also like to remind that metastability bounds depend on the fastest process conceived for the transition to the true vacuum. Any faster process occurring once anywhere in the Universe would reduce or eliminate the metastability region. 
\end{enumerate} 

We discussed in Fig.~(\ref{fig:histo}) that only a small percentage of randomly generated boundary conditions for  \(\lambda(M_{pl})\) lead to Higgs masses which are still allowed by experiments. On the other hand we presented in Fig.~(\ref{fig:pd1}) results for a set of boundary conditions which all lead to Higgs masses in the allowed region and we explained how this can be systematically understood. The point is that the chosen boundary conditions emerge from conditions which have loop suppression factors, making them rather small compared to a random choice. Fig.~(\ref{fig:pd1}) shows that there exist different working conditions and others might be found which work as well. This implies that once loop suppression factors have been included into the boundary condition it is not easily possible to distinguish between various models or scenarios, but growing precision will nevertheless reduce the number of options somewhat. 
The conclusion from this observations is in the scenario of this paper that quantum gravity should not generate a random value of \(\lambda\) at the Planck scale, but it should somehow select conditions 
that lead to a small Higgs self-coupling $\lambda$ at the Planck scale. These conditions could possibly be some remnant of symmetry in the full quantum theory of gravity. The Yukawa and \(\lambda\) couplings could or maybe even should have a common origin from the Planck scale physics, as the top quark contribution miraculously cancels the contribution of the Higgs boson such that the SM can be extrapolated to the Planck scale. 

We have shown in this paper that the chosen boundary conditions for \(\lambda\) yield a Higgs mass which is in the allowed range. The fact that different boundary conditions have overlapping regions could imply that more than one of them is simultaneously realized at the Planck scale, which is an intriguing possibility. For instance if we demand that \(\lambda(M_{pl})=0\) and \(\mathrm{Str}\mathcal{M}^2(M_{pl})=0\) are both satisfied then it is possible that the Higgs quartic coupling is only generated radiatively and the quadratic divergence vanishes. 
This appears puzzling from a low energy perspective, however, from the Planck scale physics perspective this could be natural, as these two conditions could have a common origin of some unknown connection between the gauge, Yukawa and Higgs quartic couplings. Other combinations of listed boundary conditions can also be considered, as for instance those proposed in Ref.~\cite{Shaposhnikov:2009pv} where the authors obtained from the asymptotic safety of gravity \(\lambda\approx0\) and \(\beta_{\lambda}\approx0\) near the Planck scale. Similar conditions also appeared in Ref.~\cite{Froggatt:1995rt}, where the authors demanded the principle of multiple point criticality.  

One might ask how much the situation can be improved with better measurements of the Higgs and top masses. ATLAS and CMS \cite{Hohlfeld:684112,Ball:2007zza,DeRoeck:2007zza} should be capable to detect a Higgs mass with a precision of \(0.1\%\) to \(1\%\) with an integrated luminosity of \(\unit[30]{fb}^{-1}\). 
This precision is encouraging, but unfortunately the determination of the high energy boundary conditions is plagued by the relatively large uncertainties due to the lack of higher order RGEs. Three loop beta functions and other improvements would be required in order to reduce the errors of the band in Fig.~(\ref{fig:pd1}). Without these theoretical improvements it will be hard to differentiate the different  boundary conditions. However, such improvements would be important if the SM would indeed be a theory which is valid up to the Planck scale. Precise determinations of the Higgs and top masses could then be used to identify the correct boundary conditions for \(\lambda\) within the remaining uncertainties. A complete calculation of the three-loop beta functions of the SM would be very important in this case. This would, however, also require the determination of the matching conditions to at least two-loop order. While the two-loop QCD matching condition has been included for \(\lambda_t(m_t)\) in this work, the corresponding contribution to \(\lambda(m_t)\) is not known and would be necessary to reduce the matching uncertainty alongside the other two-loop contributions to \(\lambda(m_t)\) and \(\lambda_t(m_t)\).

With around \(\unit[5]{fb^{-1}}\) of integrated luminosity collected by ATLAS and CMS, it is possible to find or exclude a SM Higgs mass in the region from \(\unit[114-600]{GeV}\) by the end of next year. A discovery of SM Higgs boson will anyhow be significant for advancement of particle physics. There exists an excellent chance to find all sort of TeV-scale new physics, but if the LHC finds nothing but a SM Higgs then this would be very much in favour of the spirit of this paper. A key question would be which new concepts could be involved in quantum gravity such that the correct SM boundary conditions arise. 

\vspace*{2mm}

\noindent {\bf Note added:} After this work has been completed, ATLAS has reported the latest Higgs mass exclusion regions at 95\% CL \cite{ATLAS-CONF-2011-163} ranging from \(\unit[112.7-115.5]{GeV}\), \(\unit[131-237]{GeV}\) and \(\unit[251-453]{GeV}\) while CMS has excluded the Higgs masses in the range \(\unit[127-600]{GeV}\) \cite{CMS-PAS-HIG-11-032}. 
\vspace*{2mm}

\noindent {\bf Acknowledgements:} We would like to thank E. Gross for interesting discussions on the exact value of the vacuum stability bound. M.H.~acknowledges support by the International Max Planck Research School for Precision Tests of Fundamental Symmetries.



\appendix
\section{List of beta functions and anomalous dimension of Higgs mass \label{sec:a1}}
We give the beta function and anomalous dimension of Higgs mass used in our calculation. The beta function for a generic coupling \(X\) is given as:
\begin{align}
\mu\frac{\mathrm{d}X}{\mathrm{d}\mu}&=\beta_X =\sum_i\frac{\beta_X^{(i)}}{(16\pi^2)^i}
\end{align}
and the anomalous dimension for Higgs mass is given as:
\begin{align}
\mu\frac{\mathrm{d}m^2}{\mathrm{d}\mu}=\gamma_m=\sum_i\frac{\gamma_m^{(i)}}{(16\pi^2)^i}
\end{align}

The list of beta functions are given below \cite{Einhorn:1992um,Luo:2002ey,Machacek:1983tz,Machacek:1983fi,Machacek:1984zw}:
\begin{widetext}
\begin{align}
\beta_{\lambda}^{(1)}&=\lambda(-9g^2_2-3g^2_1+12\lambda_t^2)+24\lambda^2+\frac{3}{4}g_2^4+\frac{3}{8}(g_1^2+g_2^2)^2-6\lambda_t^4, \\
\beta_{\lambda_t}^{(1)}&=\frac{9}{2}\lambda_t^3+\lambda_t\left(-\frac{17}{12}g_1^2-\frac{9}{4}g_2^2-8g_3^2 \right),\\
\beta_{g_1}^{(1)}&=\frac{41}{6}g_1^3,\quad \beta_{g_2}^{(1)}=-\frac{19}{6}g_2^3,\quad \beta_{g_3}^{(1)}=-7g_3^3, \\
\beta_{\lambda}^{(2)}&=-312 \lambda^3 - 144 \lambda^2 \lambda_t^2 + 36 \lambda^2 (3 g_2^2 + g_1^2) - 
3 \lambda \lambda_t^4 + \lambda \lambda_t^2 \left(80 g_3^2 + \frac{45}{2} g_2^2 + \frac{85}{6} g_1^2\right) \nonumber \\
&\quad - \frac{73}{8} \lambda g_2^4 + \frac{39}{4} \lambda g_2^2 g_1^2 + \frac{629}{24} \lambda g_1^4 + 30 \lambda_t^6 - 32 \lambda_t^4 g_3^2 - \frac{8}{3} \lambda_t^4 g_1^2 - \frac{9}{4} \lambda_t^2 g_2^4  \nonumber \\
&\quad+ \frac{21}{2} \lambda_t^2 g_2^2 g_1^2 - \frac{19}{4} \lambda_t^2 g_1^4 + \frac{305}{16} g_2^6 - \frac{289}{48} g_2^4 g_1^2 - \frac{559}{48} g_2^2 g_1^4 - \frac{379}{48} g_1^6 \\
\beta_{\lambda_t}^{(2)}&=\lambda_t \left(-12 \lambda_t^4 + \lambda_t^2 \left(\frac{131}{16} g_1^2 + \frac{225}{16} g_2^2 + 36 g_3^2 - 12 \lambda\right) + \frac{1187}{216} g_1^4 \right. \nonumber \\
&\quad \left. - \frac{3}{4} g_2^2 g_1^2 + \frac{19}{9} g_1^2 g_3^2 -\frac{23}{4} g_2^4 + 9 g_2^2 g_3^2 - 108 g_3^4 + 6 \lambda^2\right) \\
\beta_{g_1}^{(2)}&=g_1^3 \left(\frac{199}{18} g_1^2 + \frac{9}{2} g_2^2 + \frac{44}{3} g_3^2 - \frac{17}{6}\lambda_t^2\right) \\
\beta_{g_2}^{(2)}&=g_2^3 \left(\frac{3}{2} g_1^2 + \frac{35}{6} g_2^2 + 12 g_3^2 - \frac{3}{2} \lambda_t^2\right)\\
\beta_{g_3}^{(2)}&=g_3^3 \left(\frac{11}{6} g_1^2 + \frac{9}{2} g_2^2 - 26 g_3^2 -2 \lambda_t^2\right), \\
\gamma_m^{(1)}&=m^2\left(12\lambda+6\lambda_t^2-\frac{9}{2}g_2^2-\frac{3}{2}g_1^2\right) \\
\gamma_m^{(2)}&=2m^2\left(-30\lambda^2-36\lambda\lambda_t^2+12\lambda(3g_2^2+g_1^2)-\frac{27}{4}\lambda_t^4+20g_3^2\lambda_t^2\right. \nonumber \\
&\quad \left.+\frac{45}{8}g_2^2\lambda_t^2+\frac{85}{24}g_1^2\lambda_t^2-\frac{145}{32}g_2^4+\frac{15}{16}g_2^2g_1^2+\frac{157}{96}g_1^4\right)
\end{align}
\end{widetext}

\section{Matching of \texorpdfstring{$\overline{\mathrm{MS}}$}{MS-bar} coupling constant and pole mass \label{sec:a2}}
The \(\overline{\mathrm{MS}}\) gauge couplings are used in this work, therefore no matching of the gauge couplings is necessary. The boundary conditions for \(\overline{\mathrm{MS}}\) \(g_1(M_Z)\) and \(g_2(M_Z)\) couplings are taken from the value of fine structure constant \(\hat{\alpha}(M_Z)=127.916(15)\) and weak-mixing angle \(\sin^2 \hat{\theta}_W(M_Z)=0.23116(13)\) \cite{Nakamura:2010zzi} by solving the equations:
\begin{align}
\hat{\alpha}^{-1}(M_Z)&\equiv 4\pi \frac{g_1^2(M_Z)+g_2^2(M_Z)}{g_1^2(M_Z) g_2^2(M_Z)}, \\
\sin^2\hat{\theta}_W(M_Z)&\equiv \frac{g^2_1(M_Z)}{g_1^2(M_Z)+g_2^2(M_Z)}.
\end{align}
The strong coupling constant \(g_3(M_Z)\) can be extracted from \(\alpha_s(M_Z)=0.1184(7)\) \cite{Nakamura:2010zzi}.

The matching of \(\overline{\mathrm{MS}}\) top Yukawa coupling to its pole mass is given by Eq.~\eqref{eq:toppole}, where \(\delta_t\) can be split into three part \cite{Hempfling:1994ar}:
\begin{equation}
\delta_t(\mu)=\delta_{t}^{\mathrm{QCD}}(\mu)+\delta_t^{\mathrm{QED}}(\mu)+\delta_t^{\mathrm{W}}(\mu)
\end{equation}
where \(\delta_t^{\mathrm{QCD}}\) denotes the contribution of QCD correction, which we will take up to three-loop order \cite{Melnikov:2000qh}. The term \(\delta_t^{\mathrm{QED}}(\mu)+\delta_t^{\mathrm{W}}(\mu)\) contributes to the matching correction from the QED and electroweak sector. The matching terms at \(\mu=M_t\) have been calculated to be:
\begin{widetext}
\begin{align}
\delta_{t}^{\mathrm{QCD}}(M_t)&=-\frac{4\alpha_s(M_t)}{3\pi}-9.1253\left(\frac{\alpha_s(M_t)}{\pi}\right)^2-80.4046\left(\frac{\alpha_s(M_t)}{\pi}\right)^3 \\
\delta_{t}^{\mathrm{QED}}(M_t)+\delta_t^{\mathrm{W}}(M_t)&=-4\frac{\hat{\alpha}(M_t)}{9\pi}+\frac{M_t^2}{16\pi^2v^2} \left[\frac{11}{2}-\frac{M_H^2}{4M_t^2}-\frac{M_H^4}{2M_t^4}\left(\frac{4M_t^2}{M_H^2}-1\right)^{3/2}\mathrm{arccos}\left( \frac{M_H}{2M_t}\right)\right. \nonumber \\
&\quad\left. +\frac{M_H^2}{2M_t^2}\left(\frac{M_H^2}{2M_t^2}-3\right)\log\frac{M_H^2}{M_t^2}\right]-6.9\times10^{-3}+1.73\times 10^{-3}\log\frac{M_H}{\unit[300]{GeV}} \nonumber \\
&\quad-5.82\times 10^{-3}\log\frac{M_t}{\unit[175]{GeV}}
\end{align}
\end{widetext}
Note that the above formula for \(\delta_{t}^{\mathrm{QED}}+\delta_t^{\mathrm{W}}\) is only valid for \(M_H^2<4M_t^2\), which is our case in the above analysis.

As for the matching of \(\lambda\) and the Higgs pole mass given by Eq.~\eqref{eq:higgspole}, the matching correction \(\delta_H(M_t)\) is given as \cite{Sirlin:1985ux}:
\begin{align}
\displaybreak[0]
\delta_H(M_t)&=\frac{M_Z^2}{32\pi^2v^2}\left[\xi f_1(\xi)+f_0(\xi)+\xi^{-1}f_{-1}(\xi) \right] \displaybreak[0]
\end{align}
where \(\xi\equiv M_H^2/M_Z^2\) and each of the function \(f_i\) defined as:
\begin{widetext}
\begin{align}
f_1(\xi)&=6\log\frac{M_t^2}{M_H^2}+\frac{3}{2}\log\xi-\frac{1}{2}Z\left(\frac{1}{\xi} \right)-Z\left(\frac{c_w^2}{\xi}\right)-\log c_w^2+\frac{9}{2}\left(\frac{25}{9}-\frac{\pi}{\sqrt{3}} \right), \displaybreak[0]\\
f_0(\xi)&=-6\log\frac{M_t^2}{M_Z^2}\left[1+2c_w^2-2\frac{M_t^2}{M_Z^2} \right] +\frac{3c_w^2\xi}{\xi-c_w^2}\log\frac{\xi}{c_w^2}+2Z\left( \frac{1}{\xi}\right) \nonumber \displaybreak[0]\\
&\quad+\left(\frac{3c_w^2}{s_w^2}+12c_w^2 \right)\log c_w^2-\frac{15}{2}(1+2c_w^2)-3\frac{M_t^2}{M_Z^2}\left[2Z\left(\frac{M_t^2}{M_Z^2\xi}\right)\right. \nonumber \displaybreak[0]\\
&\quad\left.+4\log\frac{M_t^2}{M_Z^2}-5\right]+4c_w^2Z\left(\frac{c_w^2}{\xi} \right), \displaybreak[0]\\
f_{-1}(\xi)&=6\log\frac{M_t^2}{M_Z^2}\left[1+2c_w^4-4\frac{M_t^4}{M_Z^4} \right]-6Z\left(\frac{1}{\xi}\right)-12c_w^4Z\left(\frac{c_w^2}{\xi} \right)-12c_w^4\log c_w^2 \nonumber \displaybreak[0]\\
&\quad+24\frac{M_t^4}{M_Z^4}\left[\log\frac{M_t^2}{M_Z^2}-2+Z\left(\frac{M_t^2}{M_Z^2\xi} \right) \right]+8(1+2c_w^4).\displaybreak[0] \\
Z(z)&=\left\{
\begin{array}{l l}
2A\arctan(1/A) &\quad\mathrm{if}\, z>1/4 \\
A\log\left[(1+A)/(1-A)\right] &\quad \mathrm{if}\, z<1/4
\end{array} \quad A=\sqrt{|1-4z|}\right.
\end{align}
\end{widetext}
and \(s_w^2=\sin^2\theta_W\), \(c_w^2=\cos^2\theta_W\) and \(\theta_W\) is the Weinberg angle in on-shell scheme.

\bibliographystyle{apsrev4-1}
\bibliography{mybib}
\end{document}